# Anomalous Dome-like Superconductivity in RE$_2$(Cu$_{1-x}$Ni$_x$)$_5$As$_3$O$_2$ (RE=La, Pr, Nd)

Xu Chen,[1,2,7] Jiangang Guo,[1,7,8,*] Chunsheng Gong,[1] Erjian Cheng,[3] Congcong Le,[4,1] Ning Liu,[1,2] Tianping Ying,[3] Qinghua Zhang,[1] Jiangping Hu,[1,4,6] Shiyan Li,[3,5] and Xiaolong Chen[1,2,6,*]

[1]Beijing National Laboratory for Condensed Matter Physics, Institute of Physics, Chinese Academy of Sciences, P. O. Box 603, Beijing 100190, China
[2]University of Chinese Academy of Sciences, Beijing 100049, China
[3]State Key Laboratory of Surface Physics, Department of Physics, and Laboratory of Advanced Materials, Fudan University, Shanghai 200433, China
[4]Kavli Institute of Theoretical Sciences, University of Chinese Academy of Sciences, Beijing 100190, China
[5]Collaborative Innovation Center of Advanced Microstructures, Nanjing 210093, China
[6]Songshan Lake Materials Laboratory, Dongguan, Guangdong 523808, China
[7]These authors contributed equally
[8]Lead Contact
*Correspondence: jgguo@iphy.ac.cn (J.G.), chenx29@iphy.ac.cn (X.C.)

**SUMMARY**

**Significant manifestation of interplay of superconductivity and charge density wave, spin density wave or magnetism is dome-like variation in superconducting critical temperature ($T_c$) for cuprate, iron-based and heavy Fermion superconductors. Overall behavior is that the ordered temperature is gradually suppressed and the $T_c$ is enhanced under external control parameters. Many phenomena like pesudogap, quantum critical point and strange metal emerge in the different doping range. Exploring dome-shaped $T_c$ in new superconductors is of importance to detect emergent effects. Here, we report that the observation of superconductivity in new layered Cu-based compound RE$_2$Cu$_5$As$_3$O$_2$ (RE=La, Pr, Nd), in which the $T_c$ exhibits dome-like variation with maximum $T_c$ of 2.5 K, 1.2 K and 1.0 K as substituting Cu by large amount of Ni ions. The transitions of $T^*$ in former two compounds can be suppressed by either Ni doping or rare earth replacement. Simultaneously, the structural parameters like As-As bond length and $c/a$ ratio exhibit unusual variations as Ni-doping level goes through the optimal value. The robustness of superconductivity, up to 60% of Ni doping, reveals the unexpected impurity effect on inducing and enhancing superconductivity in this novel layered materials.**

**INTRODUCTION**

Cuprate superconductors are a class of layered compounds that belong to the regime of strongly-correlated electron system and their superconducting energy gaps are thought to be *d*-wave type (**Wollman et al., 1993; Tsuei and Kirtley, 2000**). Both magnetic and non-magnetic impurities in the Cu site will seriously suppress superconductivity (SC) (**Xiao et al., 1990**). Dome-like $T_c$ often shows up when the carrier concentration increases from 5% to 25% by doping in non-$CuO_2$ layers (spacer layers) (**Lee et al., 2006**). For iron-based superconductors, carriers change by doping in non-SC layers may have a similar effect on $T_c$ but the effect of impurities on the Fe site is totally different, which is regard as a signature of different superconducting gap symmetry details, S+- (**Mazin, 2010; Mazin et al., 2008**) or S++ (**Kuroki et al., 2008; Onari and Kontani, 2009**). For example, partial substitution of O by F in LaFeAsO (**Kamihara et al., 2008**) or Ba by K in $BaFe_2As_2$ (**Rotter et al., 2008**) can lead to the dome-like $T_c$. Surprisingly, similar dome-shaped $T_c$ also can be achieved by substituting $Fe^{2+}$ by $Co^{2+}$ or $Ni^{2+}$ ions with more 3*d* electrons (**Sefat et al., 2008; Canfield et al., 2009; Ni et al., 2010; Li et al., 2009**). One explanation is that the doping is justified by the rigid-band model to some extent, where the doped electrons are in itinerant states, only shifting the Fermi level to the higher density of states (**Ideta et al., 2013; Ideta et al., 2011**). At the same time, the correlation strength of electrons and spin fluctuations might be drastically modified (**Nakajima et al., 2014; Dai et al., 2012**). As far as we know, such dome-like $T_c$ induced by $Ni^{2+}$ has not been known in other systems.

In this work, we report three novel layered superconductors, $RE_2Cu_5As_3O_2$ (RE=La, Pr, Nd), where Cu is coordinated by As in a new kind of $[Cu_5As_3]^{2-}$ block. $La_2Cu_5As_3O_2$ (La2532) shows superconducting transition at $T_c$ = 0.63 K while the latter Pr2532 and Nd2532 are non-superconducting phases. Strikingly, dome-like $T_c$ emerges in $RE_2(Cu_{1-x}Ni_x)_5As_3O_2$ upon a wide range of Ni doping (0<*x*<0.6). These series of compounds exhibit different superconducting evolutions from both cuprate and iron-based superconductors. Our results highlight the role of Ni doping coupled with structural anomaly in inducing SC with Landau-Fermi liquid behavior.

# RESULTS

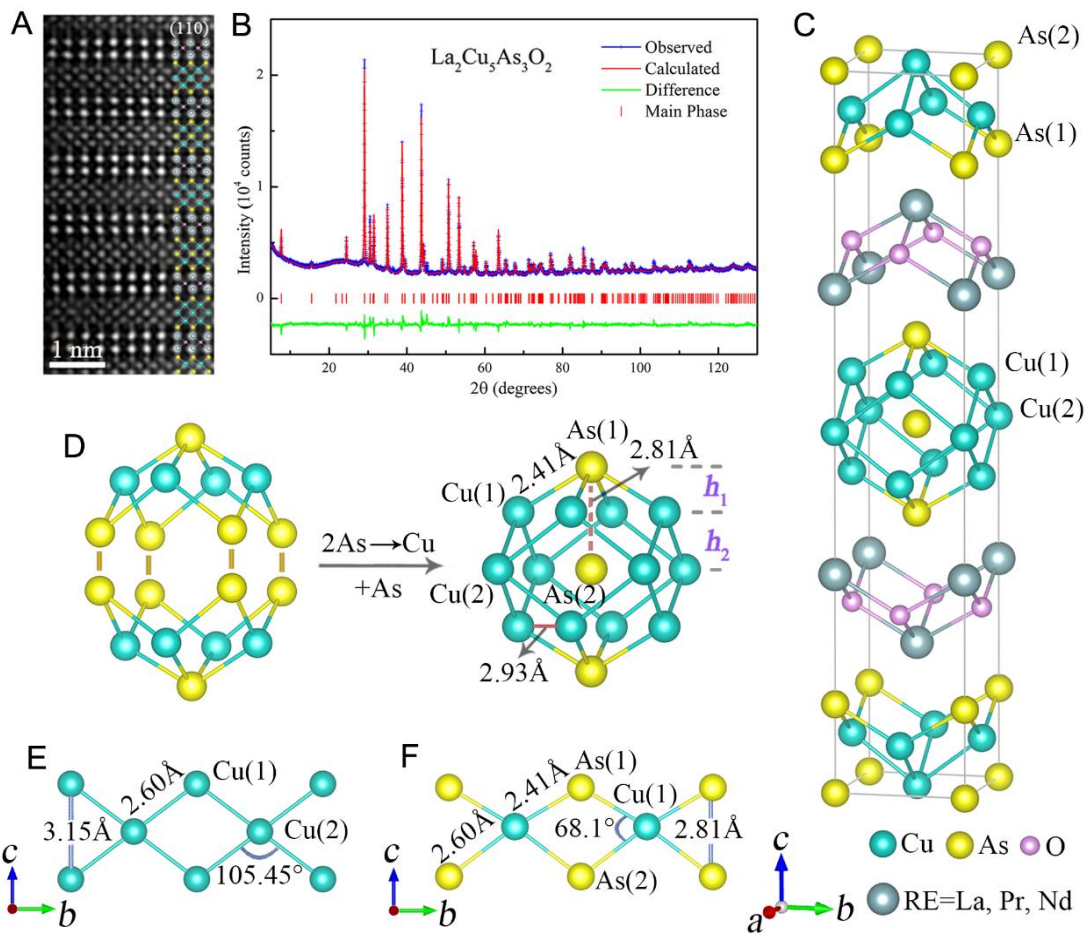

**Figure 1. Structural determination, crystal structural and bonding details of RE$_2$Cu$_5$As$_3$O$_2$.**

(A) HADDF image of (110) plane of La$_2$Cu$_5$As$_3$O$_2$.

(B) Rietveld refinement of PXRD of La2532 collected at 300 K.

(C) Crystal structure of RE$_2$Cu$_5$As$_3$O$_2$ (RE=La, Pr, Nd).

(D) [Cu$_5$As$_3$]$^{2-}$ unit is a combination of two Cu$_2$As$_2$ layers with replacing two As atoms by Cu atom. One more As atom is encapsulated in the center.

(E-F) Structural details of Cu network and Cu-As fragment.

Fig. 1A shows the HADDF image of (110) plane of La2532, in which two different slabs stack along the *c*-axis, indicating a typical layered structure. The collected powder X-ray diffraction (PXRD) pattern of La2532 can be indexed by a body-center tetragonal cell with space group *I*4/mmm (No. 139). The refined lattice constants are $a=b$=4.1386(1) Å and $c$=22.8678(6) Å. We construct the initial model by setting La1 4e (0.5, 0.5, z1), O1 4d (0.5, 0, 0.25), Cu(1) 8g (0.5, 0, z2), Cu(2) 2b (0, 0, 0.5), As(1) 4e (0, 0, z3) and As(2) 2a (0, 0, 0) as per $I4/mmm$. The Rietveld refinement successfully converges to $R_p$=2.95%, $R_{wp}$=4.26% and $\chi^2$=3.87, and the refined patterns are shown in Fig. 1B. Pr2532 and Nd2532 are found to be isostructural to La2532 with lattice parameters $a$=4.0802(1) Å and $c$=22.9144(5) Å, $a$=4.0561(1) Å and $c$=23.0082(6) Å, respectively. The crystallographic parameters of RE2532 (RE=La, Pr, Nd) are summarized in Table S1 of SM.

The crystal structure of RE2532 is drawn in Fig. 1C, one can see that the $[Cu_5As_3]^{2-}$ blocks and the fluorite $RE_2O_2$ layers stack along $c$-axis, which agrees with the atomic distributions in HADDF image and stoichiometry of EDS analysis (Fig. S1). Figure 1D is structural detail of the $[Cu_5As_3]^{2-}$ block, which can be viewed as replacing neighbor $As^{3-}$ anions of two $Cu_2As_2$ layers by one Cu atom. The bond lengths of Cu(1)-As(1) and Cu(1)-Cu(1) are 2.41 Å and 2.93 Å, respectively, close to the values in $BaCu_2As_2$ (**Saparov and Sefat, 2012**). It is noted that the bond length of As(1)-As(2), 2.81 Å, locates at the bonding regime of As-As covalent bond, 2.7~2.9 Å (**Yakita et al., 2014**). The $[Cu_5As_3]^{2-}$ unit is analogous to $[Cu_6Pn_2]^{2-}$ in $BaCu_6Pn_2$ (Pn=As, P) (**Dünner and Mewis, 1995**), where the central As atom is replaced by one Cu(2) atom. In Fig. 1, E and F, we can see that metallic bond of Cu-Cu exists in Cu network along $b$-axis as indicated by bond length, 2.60 Å, of Cu(1)-Cu(2). In coordination environment of Cu(1), short Cu(1)-As(1) and Cu(1)-As(2) bond lengths, 2.41 Å and 2.60 Å, suggest the covalence nature in $[Cu_5As_3]^{2-}$ unit like those Fe-As bonds in $[Fe_2As_2]^{2-}$ layers of iron-based superconductors (**Huang et al., 2008**).

The electrical resistivity of $RE_2Cu_5As_3O_2$ from 300 K to 1.8 K is plotted in Fig. 2, A and B. All data exhibits metallic behaviors, which can be fitted by $\rho \sim T^2$ at low temperature range, obeying the Fermi-liquid behavior, see the details in Fig. S2. There are resistivity kinks for La2532 and Pr2532 at $T^*$=80 K and 40 K, respectively. The external magnetic fields up to 9 T do not weaken this kink. However, for Nd2532, there is no any resistivity kink above 1.8 K. Measuring the resistivity at very low temperature reveals that La2532 is a superconductor with $T_c^{onset}$=0.63 K and $T_c^{zero}$=0.26 K, as shown in the inset of Fig. 2A. The transition is suppressed by external magnetic field and finally disappears as B>0.12 T. The upper critical fields $\mu_0H_{c2}(0)$, 0.15 T and 0.18 T, are estimated from the linear and Ginzburg-Landau (GL) fitting, respectively (Fig. S3). In contrast, the Pr2532 and Nd2532 are not superconductors above 0.25 K. This difference is possibly similar to the effect of suppressed SC in Pr-based cuprates (**Chen et al., 1995; Chen et al., 1992**).

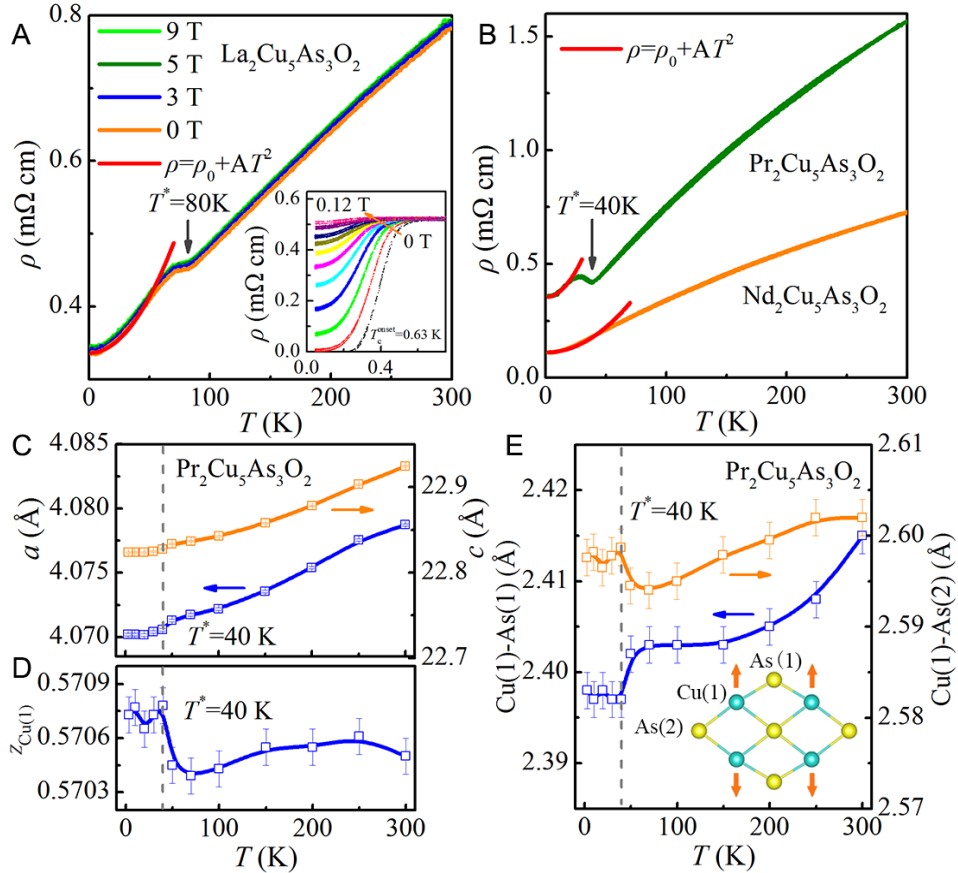

**Figure 2. Transport properties and crystal structure at various temperatures.**

(A-B) Normal-state electrical resistivity of $RE_2Cu_5As_3O_2$ as a function of temperature from 1.8 K-300 K. The data below $T^*$ can be fitted by Fermi-liquid equation. The inset is the electrical resistivity of $La_2Cu_5As_3O_2$ around $T_c$ under external magnetic field.

(C-D) Temperature-dependent $a$, $c$ and $z_{Cu(1)}$ of Pr2532.

(E) The Cu(1)-As(1) and Cu(1)-As(2) bond lengths versus temperature. Inset shows schematic variation of Cu(1) below $T^*$.

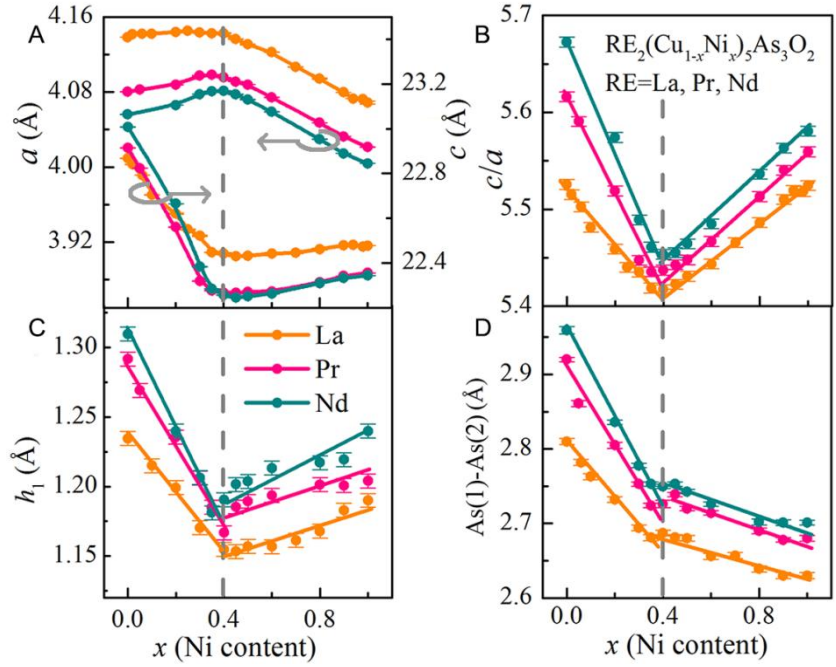

**Figure 3. Selected crystallographic parameters of $RE_2(Cu_{1-x}Ni_x)_5As_3O_2$ as a function of Ni content.**

(A) $a$ and $c$. (B) $c/a$ ratio. (C) As height ($h_1$). (D) As(1)-As(2) bond length.

The magnetic susceptibility ($\chi$) and specific heat ($C_p$) at low temperatures were measured and plotted in Fig. S4. The fitting of $\chi(T)$ gives the effective magnetic moment $\mu_{eff}$ = 0.16 $\mu_B$ and $\theta$ = -148(1) K, implying an anti-ferromagnetism (AFM) interaction of Cu ions. Furthermore, the fitting of $C_p$ yields that Debye temperature $\Theta_D$ is 169(2) K and Sommerfeld coefficient $\gamma_0$=5.01 mJ·mol$^{-1}$·K$^{-2}$ for La2532. We rule out the possibility of charge-density-wave transition by measuring the TEM of La2532 at low temperature, see Fig. S5. Meanwhile, the Rietveld refinement of temperature-dependent PXRD patterns of Pr2532 reveals that the $a$-axis and $c$-axis shrink on cooling, but both values show slightly discontinuity at 40 K, implying a structural distortion, see Fig. 2C and Fig. S6. In Fig. 2D, it is found that the $z_{Cu(1)}$ anomaly increases below 40 K. It leads to abrupt contraction of Cu(1)-As(1) bond and elongation of Cu(1)-As(2) bond, which enhances the structural anisotropy of $[Cu_5As_3]^{2-}$, as shown in Fig. 2E. Such change could induce charge redistribution and small resistivity jump. It is similar to the structural change and resistivity jump in KNi$_2$S$_2$ (**Neilson et al., 2013**).

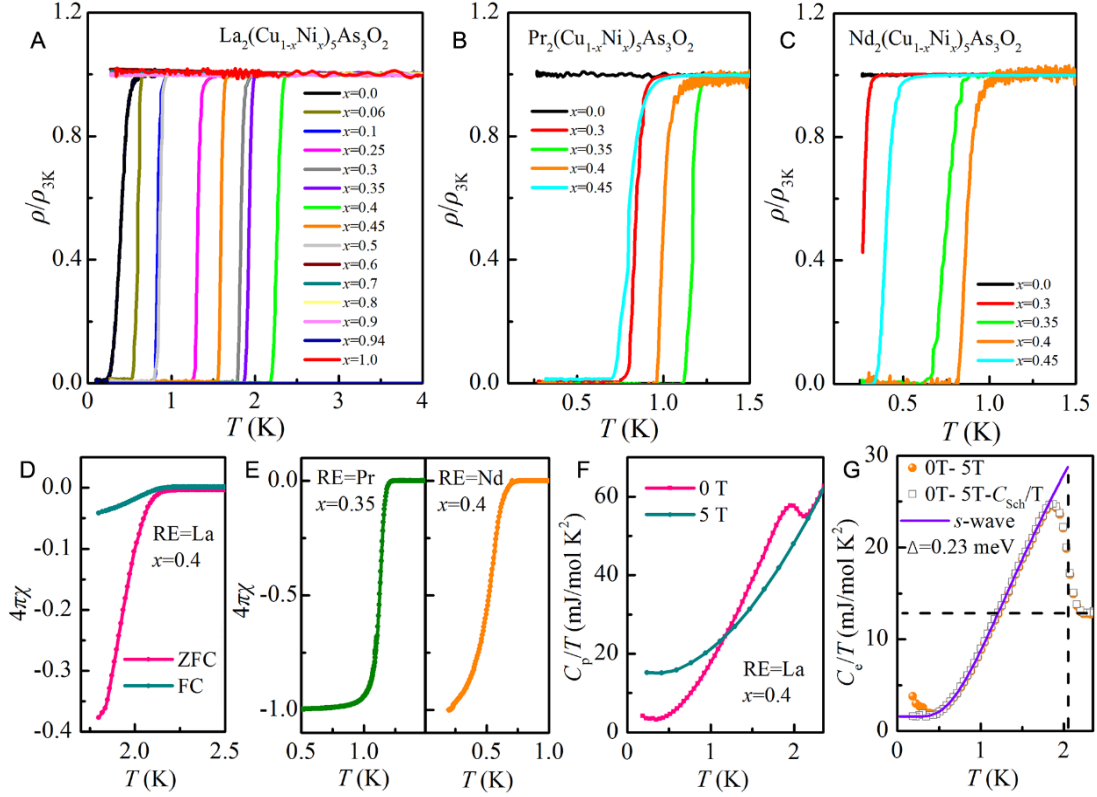

**Figure 4. Physical properties of RE$_2$(Cu$_{1-x}$Ni$_x$)$_5$As$_3$O$_2$ (RE= La, Pr, Nd).**

(A-C) Superconducting transition of RE$_2$(Cu$_{1-x}$Ni$_x$)$_5$As$_3$O$_2$. It can be found that the $T_c$ firstly increases to maximum and then decreases to zero as increasing $x$.

(D-E) Superconducting volume fraction for $x\sim0.4$ and RE=La, Pr, Nd samples at 10 Oe.

(F) $C_p/T$ of La$_2$(Cu$_{0.6}$Ni$_{0.4}$)$_5$As$_3$O$_2$ as a function of temperature under 0 T and 5 T.

(G) $C_e/T$ and the fitting curves of La$_2$(Cu$_{0.6}$Ni$_{0.4}$)$_5$As$_3$O$_2$ against temperature.

We prepared a series of RE$_2$(Cu$_{1-x}$Ni$_x$)$_5$As$_3$O$_2$ ($x$=0-1.0) samples so as to further explore the evolution SC against Ni substitution. The PXRD confirms RE$_2$(Cu$_{1-x}$Ni$_x$)$_5$As$_3$O$_2$ is a continuous solid solution, judging from the linear decrease in volume of unit cell (Fig. S7). The results of Rietveld refinement for all patterns are listed in Table S1. Selected crystallographic parameters are plotted in Fig. 3. One can see that the *c*-axis decreases drastically as $x$<0.4, but the *a*-axis almost keeps constant, however, this variation is reversed as $x$>0.4, see Fig. 3A. This anomalous feature makes the *c/a* ratio initially decreases as $x$<0.4 while it starts to increase as $x>0.4$, where the minimum shows up at $x=0.4$ shown in Fig. 3B. To our best knowledge, the structural changes of $a$, $c$ and 'V' shape of $c/a$ ratio are rare in layer superconductors. In Fig. 3C, the As height ($h_1$) firstly decreases as $x<0.4$ and then increases, where the crossover perfectly matches the structural anomaly $x=0.4$. The $h_2$ only linearly decreases upon Ni doping. The distinct variations of $h_1$ and $h_2$ induce a crossover of shrunk As(1)-As(2) bond length at $x=0.4$, see Fig. 3D. On the other hand, since the Ni-doping changes the coordination environment of Cu(1), and the Cu(2)-As(2) bond length ($\sqrt{2}*a/2$) almost keeps a constant, we speculate that the Ni firstly occupies the Cu(1) site as $x<0.4$, shortening the $h_1$ and $c$-axis.

The electrical resistivity of RE$_2$(Cu$_{1-x}$Ni$_x$)$_5$As$_3$O$_2$ at low temperature are shown in Fig. 4A-C. For La$_2$(Cu$_{1-x}$Ni$_x$)$_5$As$_3$O$_2$, the $T_c^{onset}$ monotonously increases to the maximal 2.5 K as $x=0.4$. It is surprised that SC can be induced in the non-superconducting Pr2532 and Nd2532 by Ni doping, in which the $T_c^{onset}$ also increases to the highest value 1.2 K and 1.0 K as $x\sim0.4$, respectively. Once $x$ is above 0.4, the $T_c^{onset}$ gradually decreases and finally vanishes up to 0.6. The external magnetic fields smoothly suppress the SC off, and the $\mu_0H'_{c2}(0)$ for three optimal-doped samples are 3.8 T (3.0 T), 0.69 T (0.52 T) and 0.54 T (0.37 T) estimated from the linear (GL) fitting, respectively (Fig. S8). In Fig. 4, D and E, the magnetization of three optimal-doped samples exhibit large superconducting volume fractions, indicating a bulk SC. Furthermore, the bulk SC of La$_2$(Cu$_{0.6}$Ni$_{0.4}$)$_5$As$_3$O$_2$ is confirmed by a large superconducting jump in the specific heat ($C_p$). The magnetic field up to 5 T totally suppresses the SC, as seen from Fig. 4F. We fit the $C_p$(5T) data using the equation $C_p/T=\gamma+\beta T^2$, and obtain the $\gamma=12.62$ mJ·mol$^{-1}$·K$^{-2}$, $\beta=9.89$ mJ·mol$^{-1}$·K$^{-4}$ and $\Theta_D=133(2)$ K. Extrapolating the data to 0 K finds a residual $\gamma_n$ is 1.58 mJ·mol$^{-1}$·K$^{-2}$, indicating that the non-superconducting phase is ~12.5% due to impurity. Thus we obtain the superconducting $\gamma_s$ is 11.04 mJ·mol$^{-1}$·K$^{-2}$, which results in the dimensionless jump of $C_e/\gamma_sT_c$ is 1.42, see Fig. 4G. This value is consistent with the BCS weak-coupling limit (1.43), but smaller than that of the optimal K-doped BaFe$_2$As$_2$ (2.5) (**Popovich et al., 2010**). We subtract the upturn of $C_e/T$ at very low temperature due to Schottky anomaly using the treatment in Ref.

(**Mu et al., 2007**) and obtain the flatten $C_e/T$. As the BCS theory, $C_e/T \propto \exp[-\frac{\Delta(0)}{k_B T}]$, the data are fitted, yielding superconducting gap $\Delta(0)$=0.23 meV=2.65$k_B$ K. Knowing the $\Delta(0)$, a 2$\Delta(0)/k_B T_c$ is 2.58, which is smaller than the weak-coupling limit (3.52) within the BCS framework. The estimation of the Schottky anomaly and semi-logarithmic of $C_e/T$ are shown in Fig. S9. Note that subtracting of Schottky anomaly possibly undermines the rationality of s-wave SC.

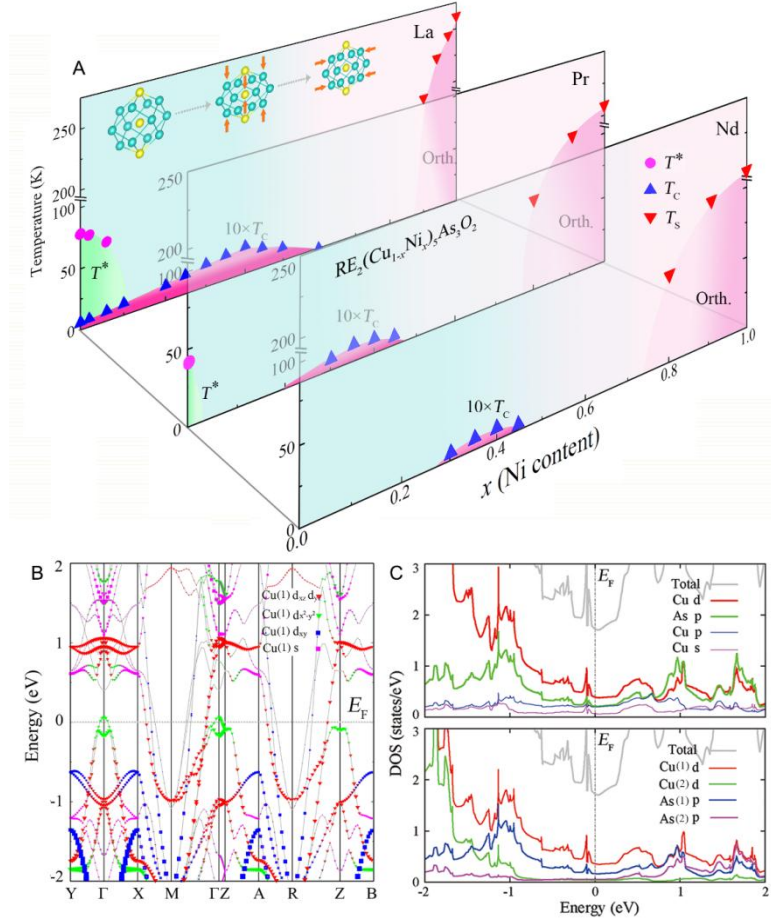

**Figure 5. Phase diagram of RE$_2$(Cu$_{1-x}$Ni$_x$)$_5$As$_3$O$_2$ and electronic structure of La$_2$Cu$_5$As$_3$O$_2$.**

(A) Phase diagram of RE$_2$(Cu$_{1-x}$Ni$_x$)$_5$As$_3$O$_2$. It can be seen that the $T^*$ is suppressed and dome-like $T_c$ shows up. The inset shows that the structure of [Cu$_5$As$_3$]$^{2-}$ unit changes upon Ni doping.

(B) Cu(1) orbital-weighted band structures of La$_2$Cu$_5$As$_3$O$_2$.

(C) Upper panel: the projected density of states of Cu $d$, $p$, $s$, and As $p$ orbitals at the ranges of -2 eV to 2 eV. Lower panel: the projected density of states of different Cu and As site, showing the Cu(1) $d$ and As(1) $p$ orbitals dominate at the Fermi level.

The electrical resistivity of RE$_2$(Cu$_{1-x}$Ni$_x$)$_5$As$_3$O$_2$ from 1.8 K-300 K shows that the $T^*$ is rapidly suppressed upon slight Ni-doping ($x$<0.1) as shown in Fig. S10. In end-member La$_2$Ni$_5$As$_3$O$_2$, Pr$_2$Ni$_5$As$_3$O$_2$ and Nd$_2$Ni$_5$As$_3$O$_2$, another resistivity anomaly associated with structural transition shows up, where the $T_s$ are 260 K, 210 K and 190 K, respectively. Indexing temperature-dependent PXRD patterns of La$_2$(Cu$_{0.02}$Ni$_{0.98}$)$_5$As$_3$O$_2$ found that the (200) and (215) peaks split into (020)/(200) and (125)/(215) peaks below $T_s$, indicating a symmetry breaking from tetragonal ($C_4$) to orthorhombic phase ($C_2$, $I$mmm, No. 71) (Fig. S11). We can map out the electronic phase diagram of RE$_2$(Cu$_{1-x}$Ni$_x$)$_5$As$_3$O$_2$, see Fig. 5A. Most interesting thing is that dome-like $T_c$ can be observed, where the superconducting phases emerge at 0<$x$<0.6, 0.2<$x$<0.45 and 0.3<$x$<0.45 for Ni-doped La2532, Pr2532 and Nd2532, respectively. The enhancement of SC upon Ni is rather rare, which is only observed in iron-based superconductors (**Sefat et al., 2008; Ni et al., 2010**). Furthermore, in terms of crystal structure, this enhancement is related to squeezing the [Cu$_5$As$_3$]$^{2-}$ along $c$-axis as $x$<0.4, and the suppression of $T_c$ corresponds to a contraction of [Cu$_5$As$_3$]$^{2-}$ along $a$-axis, shown as inset of Fig. 5A. The phase diagram is similar to those of cuprates and iron-based superconductors to large extent, which features the competition of structural distortion and SC.

**DISCUSSION**

We calculated the electronic structure in the paramagnetic state from DFT calculations. The band structures of La2532 are shown in Fig. 5B, where a small hole-pocket and three large electron-pockets show up at the Γ and M point, respectively. Around $E_F$, the bands along the Γ-X and Γ-Y directions have large dispersion while the bands along Γ-Z are almost flat, indicating that the Fermi surfaces are quasi-two-dimensional. The hole-pocket is mainly composed of Cu(1) $d_{x^2-y^2}$ hybridizing with As(1) $P_z$, and the electron-pockets components are Cu(1) $d_{xz}$/$d_{yz}$, $d_{xy}$ and As(1) $P_{x/y}$ (Fig. S12). It is noted that the Cu(1) $d_{xz}$/$d_{yz}$ dominate the states at Fermi energy ($E_F$), different from that in cuprates superconductors (**Uchida,**

**2015**). In Fig. 5C, one clearly sees that the $E_F$ is dominated by Cu(1) $d$ and As(1) $p$ states. There are higher $N(E_F)$ at -0.1 eV below $E_F$, therefore, doping Ni with one electron less can lower the $E_F$, which is theoretically reasonable to induce higher $N(E_F)$ and $T_c$. The total $N(E_F)$ is 1.75 states/eV f. u. and the estimated bare Sommerfeld coefficient is 2.06 mJ mol$^{-1}$ K$^{-2}$. According to the equation $\gamma_n=\gamma_b(1+\lambda_{e-p})(1+\lambda_{e-e})$, if we assume the electron-electron coupling $\lambda_{e-e}$=0, one would can obtain an electron-phonon coupling $\lambda_{e-p}$=1.43. The large $\lambda_{e-p}$ exceeds the limit of BCS framework, implying that the electron-electron coupling cannot be ignored.

It is previously reported that the bonded anionic dimer can induce ferromagnetic critical point, SC and metal-insulator transition (**Jia et al., 2011; Guo et al., 2012; Radaelli et al., 2002**). Here, there are weak bonding states of As(1)-As(2) in RE$_2$Cu$_5$As$_3$O$_2$, and the $E_F$ will be higher than the bonding orbital (σ), and locates the bottom of the anti-bonding orbital (σ*) (**Hoffmann and Zheng, 1985; Hoffmann, 1988**). As $x$<0.4, the doped holes would firstly enter into the As(1)-As(2) bond and lift the valence of As$^{3-}$. The strengthened bond between apical As(1)-central As(2) rapidly shortens the $c$-axis. At the same time, the $E_F$ slowly drops to the energy between σ and σ* orbital. As $x$>0.4, the shrinking of As(1)-As(2) bond length and $c$-axis slows, and then the $a$-axis begin to quickly decreases. It means that some of excess holes are introduced into the lattice, which possibly suppresses the SC. The Hall measurement, shown in Fig. S13, shows positive Hall coefficients. It indicates that the dominated carriers are holes. Using the single-band model, we obtain the carrier concentration is ~10$^{22}$ cm$^{-3}$, which is slightly increased in 40% Ni-doped La2532.

Still, the static magnetic order associated with Cu ions is not observed in all measured samples above 1.8 K, and all the $\chi(T)$ curves can be fitted by Curie-Weiss equation, see Fig. S14. The 4$f^0$ La2532 is an itinerant compound with AFM interaction ($\mu_{eff}$=0.16$\mu_B$ per Cu; θ=-148 K). For 40% and 100% Ni-doped La2532, the resultant $\mu_{eff}$ and θ are 0.56 $\mu_B$/Cu, -256 K and 0.69 $\mu_B$/Cu, -424 K, respectively. It means that the Ni-doped samples have larger $\mu_{eff}$ and stronger AFM interaction. However, these moments are still much smaller than the theoretical value (1.73 $\mu_B$) for Cu$^{2+}$ ions with $S$=1/2. It means that the magnetic interaction is not fully localized, and the emergence of Cu-Cu metallic bonds in $[Cu_5As_3]^{2-}$ significantly increases the amount of itinerant electrons. The superconductivity here is likely to be itinerant picture (**Singh and Du, 2008; Dong et al., 2008**), which is similar to those of $BaNi_2As_2$ and LaNiAsO (**Subedi and Singh, 2008; Boeri et al., 2009**). In Pr2532 and Nd2532, the total $\mu_{eff}$ and θ are 4.42 $\mu_B$/Pr and -26.5 K, 4.62 $\mu_B$/Nd and -30.4 K, respectively. The $\mu_{eff}$ is larger than the values for the magnetic $Pr^{3+}$ ($4f^2$, 3.58 $\mu_B$) and $Nd^{3+}$ ($4f^3$, 3.62 $\mu_B$) in Ni-based superconductors (**Li et al., 2014**), indicating that the magnetic contribution of Cu ions is important. Since the carrier doping can suppress the moment of $RE^{3+}$ (**Zhao et al., 2008**), we cannot summarize clear variation of magnetic moment of Cu ions in Ni-doped Pr2532 and Nd2532. High-precision experiments are called for to identify the Cu's magnetism.

The results provide a novel kind of Cu-based superconductor $RE_2Cu_5As_3O_2$ (RE=La, Pr, Nd), whose crystal structure, SC and ground states can be effectively tuned through rather wide range of Ni doping. The dome-like $T_c$ in turn is induced by the dimerization of As-As bonds along *c*-axis and shrinking of *a*-axis. The robust SC against Ni, structural anomaly and enhanced AFM interaction provides new perspectives to understand the superconducting mechanism.

**METHODS**

All methods can be found in the accompanying Transparent Methods supplemental file.

**SUPPLEMENTAL INFORMATION**

Supplemental Information includes Transparent Methods, 14 figures, and 1 tables

**ACKNOWLEDGMENTS**

We acknowledge H. Hosono, H. Ding and Y. Zhang for valuable discussions and TEM measurement. This work was supported by the National Key Research and Development Program of China (No. 2017YFA0304700, 2016YFA0300600), the National Natural Science Foundation of China (No. 51772322, 51532010), and the Strategic Priority Research Program and Key Research Program of Frontier Sciences of the Chinese Academy of Sciences (No. QYZDJ-SSW-SLH013).

**AUTHOR CONTRIBUTIONS**

J. G and X. L. C. provided strategy and advice for the material exploration. X. C., J. G. and C. Gong performed the sample fabrication, measurements and fundamental data analysis. E. C., T. Y. and S. L. measured the low temperature properties. C. L., N. L. and J. H. carried out the theoretical calculation. Q. Z. measured the HAADF images. J. G. and X. L. C. wrote the manuscript based on discussion with all the authors.

**DECLARATION OF INTERESTS**

The authors declare no competing interests.